\PassOptionsToPackage{unicode}{hyperref}
\PassOptionsToPackage{hyphens}{url}
\documentclass[
]{article}
\usepackage{xcolor}
\usepackage{amsmath,amssymb}
\usepackage{iftex}
\usepackage{graphicx}
\ifPDFTeX
  \usepackage[T1]{fontenc}
  \usepackage[utf8]{inputenc}
  \usepackage{textcomp} 
\else 
  \usepackage{unicode-math} 
  \defaultfontfeatures{Scale=MatchLowercase}
  \defaultfontfeatures[\rmfamily]{Ligatures=TeX,Scale=1}
\fi
\usepackage{lmodern}
\ifPDFTeX\else
\fi
\IfFileExists{upquote.sty}{\usepackage{upquote}}{}
\IfFileExists{microtype.sty}{
  \usepackage[]{microtype}
  \UseMicrotypeSet[protrusion]{basicmath} 
}{}
\makeatletter
\@ifundefined{KOMAClassName}{
  \IfFileExists{parskip.sty}{%
    \usepackage{parskip}
  }{
    \setlength{\parindent}{0pt}
    \setlength{\parskip}{6pt plus 2pt minus 1pt}}
}{
  \KOMAoptions{parskip=half}}
\makeatother
\usepackage{longtable,booktabs,array}
\usepackage{calc} 
\usepackage{etoolbox}
\makeatletter
\patchcmd\longtable{\par}{\if@noskipsec\mbox{}\fi\par}{}{}
\makeatother
\IfFileExists{footnotehyper.sty}{\usepackage{footnotehyper}}{\usepackage{footnote}}
\makesavenoteenv{longtable}
\providecommand{\tightlist}{%
  \setlength{\itemsep}{0pt}\setlength{\parskip}{0pt}}
\usepackage{bookmark}
\IfFileExists{xurl.sty}{\usepackage{xurl}}{} 
\hypersetup{
  hidelinks,
  pdfcreator={LaTeX via pandoc}}

\author{}
\date{}

\begin{document}

\section{PHBench: A Benchmark for Predicting Startup Series A Funding from Product Hunt Launch Signals}\label{phbench-a-benchmark-for-predicting-startup-series-a-funding-from-product-hunt-launch-signals}

Yagiz Ihlamur\textsuperscript{1}, Ben Griffin\textsuperscript{2}, Rick Chen\textsuperscript{2}

\textsuperscript{1} Amazon, Seattle, WA, USA \textsuperscript{2} University of Oxford, Oxford, UK

This work was conducted independently by the authors and does not represent the views or projects of Amazon.

Correspondence: ihlamuryagiz@gmail.com

Website: https://phbench.com

Dataset: Available on request (anonymized public splits)

\begin{center}\rule{0.5\linewidth}{0.5pt}\end{center}

\subsection{Abstract}\label{abstract}

Structured launch signals on Product Hunt contain statistically significant predictive information for Series A funding outcomes. We construct PHBench from 67,292 featured Product Hunt posts spanning 2019--2025, linked to Crunchbase funding records via deterministic domain matching, identifying 528 verified Series A raises within 18 months of launch (positive rate: 0.78\%). Our best-performing model, a three-component ensemble (ENS\_avg, ENS\_ISO, XGB) selected by validation $F_{0.5}$, achieves $F_{0.5}$ = 0.097 and AP = 0.037 (95\% CI: 0.024--0.072; 4.7$\times$ lift over random) on the private held-out test set (103 positives). A paired bootstrap confirms a statistically credible advantage over the logistic regression baseline (AP delta: +0.013, 95\% CI: {[}0.004, 0.039{]}, p \textless{} 0.001; $F_{0.5}$ delta: +0.056, 95\% CI: {[}0.006, 0.122{]}, p = 0.016). Validation-set metrics ($F_{0.5}$ = 0.284, AP = 0.126) reflect best-of-144 selection bias on 53 positives and are reported for benchmark reproducibility only.

We further evaluate three zero-shot Gemini models (Gemini 2.5 Flash, Gemini 3 Flash, and Gemini 3.1 Pro) in an anonymized numerical setting. The best LLM achieves AP = 0.034 (Gemini 3 Flash), below the LR baseline AP of 0.044. Notably, the most capable Gemini variant (Gemini 3.1 Pro, AP = 0.023) performs worst --- an unexpected pattern that warrants further investigation across providers and prompting strategies. Both ML and LLM models show the same temporal performance decay tracking the 2020--2021 funding boom and subsequent contraction, confirming the dataset captures genuine market structure rather than noise.

PHBench provides a reproducible framework comprising public training, validation, and blind test splits; 61 engineered features; a five-metric evaluation harness; and a public leaderboard at phbench.com. All code, baseline models, and anonymized dataset splits are publicly available.

\subsection{1. Introduction}\label{introduction}

In 2023, only 0.78\% of Product Hunt featured launches led to a Series A funding round within 18 months, yet the launches that did shared measurable patterns: higher upvote counts, stronger daily rankings, larger maker teams, and a tilt toward B2B market categories. Early-stage investors evaluate thousands of companies annually, yet the signals that distinguish eventual Series A raises from the broader population remain poorly understood and largely unformalized. Analysts rely on pattern recognition honed over years of experience, but this knowledge is neither transferable nor reproducible. A quantitative benchmark grounded in observable pre-funding signals would make the evaluation of predictive models tractable and allow systematic comparison across methods.

Product Hunt (PH) provides a structured community validation mechanism particularly well-suited to this problem. Each product launch receives a curated 24-hour window in which the community votes, comments, and signals engagement. Founders strategically coordinate launches; investors monitor the platform. The resulting launch record (upvotes, comments, rankings, maker profiles, and topic tags) constitutes a rich snapshot of early market reception, captured at a single point in time and indexed by a precise timestamp. Critically, this metadata is archived in a queryable API, making large-scale retrospective collection feasible.

We define the prediction task as follows: given the launch-day metadata of a Product Hunt post, predict whether the associated company will raise a Series A round within 18 months of the launch date. This formulation is practically relevant: a model that ranks launches by funding probability enables weekly screening of the PH feed, reducing analyst workload while surfacing high-signal opportunities. The 18-month window is chosen to capture the fundraising cycle that typically follows a successful public launch, while excluding companies already funded before launch. Although the median launch-to-Series-A interval is 265 days (8.7 months), a meaningful long tail of raises extends well beyond 12 months, a shorter window would systematically exclude them. We provide a full distribution histogram in Appendix C.

This paper makes three contributions. First, we release a curated dataset of 67,292 featured Product Hunt posts linked to 528 verified Series A outcomes via deterministic domain matching against a Crunchbase snapshot of 374,150 companies, alongside the linkage methodology and a 50-domain platform blocklist used to suppress false attribution. Second, we establish PHBench as an evaluation framework over this dataset: stratified train/validation/test splits with company-level grouping, 61 reproducible engineered features, a five-metric harness, private test labels held by the maintainers and scored server-side, and a public leaderboard at phbench.com. Third, we provide a systematic comparison of gradient-boosted ensemble models against three frontier LLMs evaluated under an anonymized zero-shot protocol, establishing the first quantitative LLM baseline on this task.

The paper proceeds as follows. Section 2 reviews related work on startup success prediction, social proof signals, and ML benchmarks. Section 3 describes dataset construction, including data collection, Crunchbase linkage, label construction, and known limitations. Section 4 defines the benchmark design, split methodology, and evaluation metrics. Section 5 details the 61 engineered features and their design rationale. Section 6 reports results across 144 ML experiments with analysis of the champion ensemble. Section 7 presents LLM results and a four-finding analysis. Section 8 discusses implications and concludes.

\begin{center}\rule{0.5\linewidth}{0.5pt}\end{center}

\subsection{2. Related Work}\label{related-work}

\subsubsection{2.1 Startup Success Prediction}\label{startup-success-prediction}

A growing body of literature applies machine learning to predict startup outcomes. Arroyo et al. {[}Arroyo2019{]} evaluate ML models for VC investment decision support using Crunchbase data, applying time-aware multi-class prediction across company, funding, and founder features to identify the most promising early-stage investments. Gastaud et al. {[}Gastaud2019{]} examine multiple funding stages and demonstrate that feature importance shifts significantly across rounds, with seed-stage predictions relying heavily on team signals while later-stage predictions incorporate market traction. More recently, Griffin et al. {[}Griffin2025{]} propose Random Rule Forest, an interpretable ensemble of LLM-generated YES/NO questions evaluated against founder profiles, demonstrating that LLM-generated natural-language questions can serve as predictive features for early-stage investment screening. These studies confirm the statistical predictability of funding events from both structured metadata and LLM-derived features, but operate on proprietary or single-snapshot datasets that cannot be extended or independently validated. Chen et al. {[}Chen2025{]} partially address this gap with VCBench, a public benchmark for founder-profile prediction with held-out evaluation; PHBench complements this by shifting from founder signals to product launch signals.

\subsubsection{2.2 Social Proof and Community Signals}\label{social-proof-and-community-signals}

Community engagement signals carry predictive information in adjacent domains. Social proof mechanisms, where early adoption signals quality to subsequent observers, are well-documented in crowdfunding research: Mollick {[}Mollick2014{]} finds that early momentum in crowdfunding campaigns is strongly predictive of ultimate success, with projects that achieve early traction attracting disproportionate subsequent support. This herding dynamic has direct analogues on Product Hunt, where a product that achieves high early engagement gains algorithmic visibility, attracting further votes and comments in a self-reinforcing cycle. PHBench operationalizes this mechanism at launch time: we treat the 24-hour launch record as a compressed observation of early social proof, and test whether this signal predicts downstream institutional funding outcomes 18 months later.

\subsubsection{2.3 ML Benchmark Datasets}\label{ml-benchmark-datasets}

The machine learning community has progressively standardized evaluation through curated benchmarks. Vanschoren et al. {[}Vanschoren2014{]} introduced OpenML as a shared platform for reproducible ML experiments across diverse tasks. Jimenez et al. {[}Jimenez2024{]} demonstrate with SWE-bench that carefully constructed held-out evaluation sets (where test labels are never released) prevent contamination and enable meaningful progress measurement across model generations. PHBench adopts this design: test labels remain private, and all submissions are scored by the benchmark maintainers against held-out data.

Most relevant to PHBench is VCBench {[}Chen2025{]}, which benchmarks frontier LLMs on founder success prediction using 9,000 anonymized founder profiles. VCBench finds that DeepSeek-V3 achieves 6$\times$ the market baseline precision, and most evaluated LLMs outperform human expert benchmarks (YC at 1.7$\times$ and tier-1 firms at 2.9$\times$ the baseline), demonstrating that LLMs can meaningfully outperform expert human judgment when given rich semantic founder data. PHBench complements VCBench by shifting from founder career signals to product launch signals, and by introducing a direct ML-vs-LLM comparison on the same prediction task. A related benchmark {[}Ihlamur2026{]} evaluates frontier LLMs on structured VC reasoning using career trajectory data, finding systematic performance gaps on domain-specific prediction.

\subsubsection{2.4 LLMs on Tabular Data}\label{llms-on-tabular-data}

The performance of LLMs on tabular prediction tasks has emerged as an active research question. Hegselmann et al. {[}Hegselmann2023{]} (TabLLM) demonstrate that LLMs can leverage textual metadata to improve predictions on structured medical datasets, outperforming gradient-boosted models when feature semantics are informative. Hollmann et al. {[}Hollmann2023{]} (TabPFN) show that prior-data-fitted neural networks achieve competitive performance on small tabular benchmarks, raising questions about when structured ML retains its advantage.

PHBench contributes a data point on the opposite side: our LLM evaluation uses anonymized numerical signals only, deliberately excluding semantic content (product names, taglines, descriptions). This isolates whether LLMs possess implicit knowledge of funding dynamics or whether they require semantic context to perform. The results suggest that in the purely numerical regime, gradient-boosted models substantially outperform zero-shot LLMs, consistent with the general finding that LLMs underperform on tabular tasks without strong semantic features {[}Hollmann2023{]}.

\begin{center}\rule{0.5\linewidth}{0.5pt}\end{center}

\subsection{3. Dataset Construction}\label{dataset-construction}

\subsubsection{3.1 Data Collection}\label{data-collection}

We collected Product Hunt posts via the PH GraphQL API v2, paginating through all posts using postedAfter / postedBefore parameters with a rate limit of 6,250 complexity points per 15-minute rolling window. Collection spans Q1 2019 through Q4 2025, yielding 288,692 total posts. We restrict the benchmark universe to \emph{featured} posts (those with a non-null featuredAt timestamp), which represents the curated daily front page and excludes posts that failed to reach the editorial threshold. This yields 67,292 featured posts.

For each post we collect: post ID, name, slug, tagline, description (capped at 2,000 characters), vote count, comment count, review count, review rating, creation and feature timestamps, the raw PH redirect URL, pipe-separated topic names, per-maker fields (ID, name, follower count), and rank signals (daily, weekly, monthly, and yearly rank) via the PH GraphQL API.

Website URL resolution requires special handling. We resolve real destination URLs via automated browser-compatible requests, achieving high resolution coverage across featured posts.

\subsubsection{3.2 Crunchbase Linkage}\label{crunchbase-linkage}

We obtained a Crunchbase snapshot (April 2026) containing 374,150 for-profit companies with associated funding records. Linkage to Product Hunt posts proceeds via domain matching: we normalize both the resolved PH website domain and the CB homepage\_url domain by stripping https://, www., and trailing slashes, then lowercasing.

Platform and hosting domains introduce systematic false attribution. For example, naive matching assigns 886 Product Hunt posts to hugedomains.com and hundreds more to github.com, apps.apple.com, webflow.io, and similar shared infrastructure. We construct a blocklist of 50 platform domains that are excluded from matching. These are identified by examining the top match targets by post count and manually classifying each as platform/hosting or legitimate company domain.

The resulting match rates are 15,781 matches across all 288,692 posts (5.47\%) and 8,275 matches in the featured subset (12.3\%). The higher match rate in the featured subset reflects the editorial filtering: featured products are more likely to have an established web presence and to have attracted institutional funding.

\subsubsection{3.3 Label Construction}\label{label-construction}

A post is labeled positive if: (1) the matched Crunchbase record has a confirmed Series A raise (cb\_raised\_series\_a = 1), and (2) the first Series A announcement date falls between 0 and 548 days after the Product Hunt launch date. The 548-day threshold corresponds to 18 months (365 + 183 days), providing a precise and reproducible window definition. The 18-month window is approximately twice the median observed launch-to-Series-A interval (265 days), chosen to capture the long tail of the funding distribution while remaining within a practically meaningful investment horizon. The strictly positive lower bound (0 days) excludes companies that used Product Hunt as a go-to-market channel after already raising Series A, ensuring that labels reflect \emph{predictive} rather than \emph{contemporaneous} funding.

This procedure identifies 528 positive labels out of 67,292 featured posts, yielding a class imbalance ratio of 1:126 (positive-to-negative) and a positive rate of 0.78\%. Note that this rate is computed at the post level, not the company level. Because a company can appear multiple times on Product Hunt across different launches, the post-level positive rate is not directly comparable to a real-world company-level Series-A-within-18-months base rate. This imbalance is representative of realistic deal-flow screening and is a deliberate property of the benchmark: models must be evaluated on precision-weighted metrics rather than accuracy.

The benchmark spans Q1 2019 through Q4 2025, reflecting a Crunchbase snapshot from April 2026. Posts from 2025 launches are included but form a partially-resolved cohort: their 18-month prediction window extends to mid-2026 and beyond, so some current negative labels may convert as future rounds are announced. Based on historical conversion rates (\textasciitilde0.78\%), we estimate fewer than 35 affected posts, under 0.07\% of the benchmark universe. All reported results treat the April 2026 snapshot as ground truth.

\subsubsection{3.4 Dataset Characterization}\label{dataset-characterization}

Table 1 reports year-over-year conversion rates on the combined training and validation sets for 2019--2024. The 2025 cohort (4,076 posts) is excluded from the year-level breakdown: the 18-month Series A window is not fully resolved for launches after mid-2024, making 2025 labels systematically incomplete.

\textbf{Table 1: Year-over-year Series A conversion rates (train + val combined, 2019--2024)}

{\def\LTcaptype{none} 
\begin{longtable}[]{@{}lrrrr@{}}
\toprule\noalign{}
Year & Posts & Series A & Rate & vs. Base Rate \\
\midrule\noalign{}
\endhead
\bottomrule\noalign{}
\endlastfoot
2019 & 6,122 & 48 & 0.78\% & 0.98$\times$ \\
2020 & 7,264 & 94 & \textbf{1.29\%} & \textbf{1.62$\times$} \\
2021 & 10,097 & 108 & \textbf{1.07\%} & \textbf{1.34$\times$} \\
2022 & 8,448 & 42 & 0.50\% & 0.62$\times$ \\
2023 & 11,991 & 66 & 0.55\% & 0.69$\times$ \\
2024 & 5,826 & 40 & 0.69\% & 0.86$\times$ \\
\textbf{All (2019--2024)} & \textbf{49,748} & \textbf{398} & \textbf{0.80\%} & --- \\
\end{longtable}
}

The 2020--2021 peak in conversion rate (1.29\% and 1.07\% respectively) reflects the broader venture capital funding boom of that period, where the 18-month prediction window captures the 2021--2022 Series A surge. The sharp decline in 2022--2023 cohort conversion (0.50--0.55\%) corresponds to the well-documented VC funding contraction of that period {[}PitchBook2023{]}, during which global venture investment fell sharply from its 2021 peak. The 2024 cohort shows partial recovery (0.69\%). This temporal variation is an intentional feature of the benchmark: models that achieve high performance must generalize across market cycles, not merely exploit a single funding regime.

Among the 528 verified positives, 88.3\% had prior seed funding before their Series A. The median time from PH launch to Series A announcement is 265 days (approximately 8.7 months). The median Series A round size is \$15M (mean \$21.7M), consistent with institutional early-stage financing in technology markets. A notable example in the dataset is Lovable, which launched on Product Hunt with 1,477 votes, daily rank \#1, and subsequently raised a \$200M Series A 237 days after launch\textsuperscript{2} {[}Lovable2025{]}.

Topic-level conversion rates reveal strong market segment effects. Among topics with at least 50 posts, the five highest-converting are: API (2.98\%), Payments (2.42\%), Fintech (2.41\%), Meetings (2.31\%), and Sales (2.15\%), all B2B categories, averaging approximately 3$\times$ the baseline conversion rate.

\subsubsection{3.5 Limitations}\label{limitations}

\textbf{Crunchbase coverage bias.} Crunchbase is systematically more complete for US-based, institutionally-backed companies. International startups, bootstrapped companies, and those that raise from non-institutional angels are underrepresented, introducing geographic and financing-style bias into the positive labels. Independent of these systematic biases, Crunchbase itself has incomplete coverage even within its target population --- funding rounds and entire companies can be absent from any given snapshot, contributing additional false negatives beyond those introduced by the domain-matching step.

\textbf{Domain matching conservatism.} The domain-matching approach is intentionally conservative: approximately 80\% of posts with live websites have no corresponding Crunchbase record. This produces false negatives (funded companies without a match are labeled 0) but avoids the precision cost of approximate matching methods.

\textbf{Announcement vs. close date.} Crunchbase Series A dates reflect public announcement, not the legal close of the round. The true close typically precedes announcement by one to three months, meaning the effective prediction window is shifted earlier by a similar amount rather than spanning exactly 0--18 months post-launch.

\textbf{Maker follower temporal leakage.} Maker follower counts (maker\_followers\_total, log\_maker\_followers) were scraped during data collection in 2026, not at the time of the original launch. These features therefore reflect post-hoc follower growth and introduce a subtle form of temporal leakage for historical posts. An ablation removing these two features shows an AP reduction of 0.008 on the validation set, within bootstrap sampling variance at 53 positives and best treated as indicative. We retain them with this documented limitation; the signal will attenuate for future-cohort evaluations as the scrape-time gap closes. This leakage is asymmetric across cohorts: 2019 launches have approximately seven years of post-launch follower accumulation captured in the 2026 scrape, while 2024--2025 launches have months. This asymmetry may contribute to the AP decay observed across cohorts in Finding 4 (Section 7.5), beyond the base-rate variation documented there. A full test-set and per-cohort ablation of maker follower features is left to future work.

\subsubsection{3.6 Anonymization}\label{anonymization}

Public splits use a salted SHA-256 hash to map Product Hunt post IDs to anonymous identifiers (anon\_id). The following fields are stripped from public releases: product name, slug, tagline, description, resolved website URL, maker names, all Crunchbase fields, and the raw days\_to\_series\_a value. Retained features include all 61 engineered numerical and categorical features, engagement counts, rank values, topic flags, and temporal bucket assignments. The private ID mapping linking anon\_id to real post IDs is held by the benchmark maintainers and is never released.

\begin{center}\rule{0.5\linewidth}{0.5pt}\end{center}

\subsection{4. Benchmark Design}\label{benchmark-design}

\subsubsection{4.1 Split Methodology}\label{split-methodology}

The dataset is partitioned into training (70\%), validation (10\%), and test (20\%) splits. We allocate 20\% to test to ensure sufficient positives for stable evaluation by future participants (whose submissions are scored against held-out test labels), and 10\% to validation for hyperparameter selection and reporting baseline results. Stratification is performed jointly on year\_bucket $\times$ topic\_bucket $\times$ label, ensuring that the positive rate and temporal/topical distributions are approximately preserved across all three splits.

Critically, company-level grouping is enforced: all posts associated with the same Crunchbase permalink are assigned to a single split. This prevents label leakage through product variants or re-launches from the same company. We verify this constraint by confirming zero overlapping cb\_permalink values across splits. The final split sizes are:

\begin{itemize}
\item
  Training set: 47,071 posts, 372 positives (0.79\%)
\item
  Validation set: 6,753 posts, 53 positives (0.78\%)
\item
  Test set: 13,468 posts, labels withheld
\item
  Total (public): 53,824 posts, 425 positives (0.79\%)
\end{itemize}

\subsubsection{4.2 Evaluation Metrics}\label{evaluation-metrics}

\textbf{Primary metric: $F_{0.5}$ score.} The $F_{0.5}$ score is the harmonic mean of precision and recall with $\beta$ = 0.5, weighting precision twice as heavily as recall. The rationale is practical: in VC deal-flow screening, false positives (analyst time spent on non-qualifying companies) are more costly than false negatives (missed deals) because the former consume scarce capacity while the latter are recoverable through other sourcing channels. $F_{0.5}$ is used as the primary leaderboard ranking metric at phbench.com, penalizing models that flood the analyst queue with false leads.

\textbf{Secondary metric: Average Precision (AP).} AP computes the area under the precision-recall curve, providing a threshold-free summary of ranking quality across the full operating range. AP is the standard metric for highly imbalanced classification tasks and is more stable than AUC-ROC when positive rates are very low {[}Davis2006{]}.\textsuperscript{1}

\textbf{Precision@k (P@50, P@100, P@200).} These metrics measure precision among the top-k highest-probability predictions. They are operationally meaningful: a weekly PH feed produces approximately 50--200 featured posts, and an analyst team reviewing top-k predictions must achieve high precision to remain useful.

\textbf{AUC-ROC.} Area under the receiver operating characteristic curve, a standard discriminative performance metric, included for comparability with prior startup prediction literature.

\subsubsection{4.3 Statistical Considerations}\label{statistical-considerations}

\textbf{Table 2: Dataset split statistics}

{\def\LTcaptype{none} 
\begin{longtable}[]{@{}lrrrl@{}}
\toprule\noalign{}
Split & Posts & Positives & Positive Rate & Labels \\
\midrule\noalign{}
\endhead
\bottomrule\noalign{}
\endlastfoot
Training & 47,071 & 372 & 0.79\% & Public \\
Validation & 6,753 & 53 & 0.78\% & Public \\
Test & 13,468 & 103$^\dagger$ & 0.76\% & Withheld \\
\textbf{Total (public)} & \textbf{53,824} & \textbf{425} & \textbf{0.79\%} & --- \\
\end{longtable}
}
{\footnotesize{$^\dagger$ Test set positive count disclosed by benchmark maintainers for calibration purposes; individual labels remain withheld from participants.}}

The private test set contains 103 positives (0.76\% positive rate). The small number of validation positives (53) has a significant implication for metric reliability. All Precision@k estimates are subject to substantial variance at this scale: a difference of one correct prediction changes P@50 by 2 percentage points. Researchers reporting validation-set results should report confidence intervals wherever possible. We recommend bootstrap resampling (B = 1,000) over the positive pool as the standard approach for constructing 95\% CIs for P@k and AP on this benchmark.

The class imbalance ratio of 1:126 (positive-to-negative) is an intentional property of the dataset. Models trained or evaluated on this benchmark must use imbalance-aware metrics (AP, $F_{0.5}$) as primary signals; accuracy-based metrics are uninformative at this ratio.

\subsubsection{4.4 Submission Protocol}\label{submission-protocol}

Participants receive public training, validation, and test feature files (with anonymized anon\_id identifiers) upon request. Submissions consist of a CSV file with columns anon\_id and series\_a\_prob, containing probability predictions for all 13,468 test posts. The benchmark maintainer translates anon\_id values via a private mapping file and scores predictions against held-out test labels. Results are posted to the public leaderboard at phbench.com. Dataset files (phbench\_public\_train.csv, phbench\_public\_validation.csv, phbench\_public\_test.csv, and a sample submission template) are available on request by emailing benchmark@vela.partners with subject line `PHBench Data Access'.

\begin{center}\rule{0.5\linewidth}{0.5pt}\end{center}

\subsection{5. Feature Engineering}\label{feature-engineering}

\subsubsection{5.1 Feature Groups}\label{feature-groups}

We engineer 61 features across seven groups from raw launch-day metadata. Table 3 summarizes the feature groups.

\textbf{Table 3: Feature group summary}

{\def\LTcaptype{none} 
\begin{longtable}[]{@{}
  >{\raggedright\arraybackslash}p{(\linewidth - 4\tabcolsep) * \real{0.3333}}
  >{\raggedleft\arraybackslash}p{(\linewidth - 4\tabcolsep) * \real{0.3333}}
  >{\raggedright\arraybackslash}p{(\linewidth - 4\tabcolsep) * \real{0.3333}}@{}}
\toprule\noalign{}
\begin{minipage}[b]{\linewidth}\raggedright
Group
\end{minipage} & \begin{minipage}[b]{\linewidth}\raggedleft
Count
\end{minipage} & \begin{minipage}[b]{\linewidth}\raggedright
Key Features
\end{minipage} \\
\midrule\noalign{}
\endhead
\bottomrule\noalign{}
\endlastfoot
Engagement & 6 & log\_votes, log\_comments, log\_reviews, votes\_per\_comment, has\_reviews, review\_rating \\
Rank & 8 & daily\_rank, weekly\_rank, monthly\_rank, log\_daily\_rank, log\_weekly\_rank, rank\_bucket (ordinal 0--5), has\_daily\_rank, has\_weekly\_rank \\
Maker & 4 & maker\_count, log\_maker\_followers, maker\_followers\_total, has\_makers \\
Temporal & 7 & launch\_dow, is\_prime\_window, launch\_year, launch\_quarter, launch\_month, days\_since\_ph\_launch, is\_weekday \\
Topics & 24 & 20 binary topic flags + topic\_count + is\_ai\_topic + is\_b2b\_topic + is\_consumer\_topic \\
Text & 8 & tagline\_length, tagline\_word\_count, desc\_length, desc\_word\_count, has\_description, tagline\_has\_number, tagline\_has\_percent, tagline\_has\_dollar \\
Interactions & 4 & makers\_x\_votes, votes\_x\_rank\_bucket, prime\_window\_x\_votes, ai\_topic\_x\_year \\
\textbf{Total (used)} & \textbf{61 used in reported experiments (62 released --- maker\_tier is included in the public dataset CSV but excluded from all reported models; see Section 5.2)} & \\
\end{longtable}
}

\subsubsection{5.2 Key Design Decisions}\label{key-design-decisions}

\textbf{Log transformations.} Engagement counts (votes, comments, reviews, maker followers) follow heavily right-skewed distributions. We apply log(1 + x) transformations to all count features, which improves linear model performance and reduces the influence of extreme outliers.

\textbf{Rank bucket encoding.} The rank\_bucket feature encodes daily rank into an ordinal scale: 0 = unranked, 1 = rank \#11 or below, 2 = \#6--10, 3 = \#4--5, 4 = \#2--3, 5 = \#1. This captures the non-linear relationship between rank and conversion: the Mann-Whitney U test confirms that ranked posts significantly outperform unranked posts (p = 2.16$\times$10$^{-}$\textsuperscript{5}\textsuperscript{3} for weeklyRank), and within ranked posts, rank \#1 achieves a conversion rate of 2.22\% versus 0.51\% for rank \#11 and below.

\textbf{Maker tier removal.} An initial feature set of 62 features included a maker\_tier variable encoding the cumulative follower-count tier of the maker team. This feature showed zero gain importance across all tree-based models in a preliminary sweep and, upon ablation, its removal increased Average Precision by 0.016. We attribute this to collinearity with maker\_count and log\_maker\_followers, causing maker\_tier to act as noise. The feature is excluded from all reported experiments. maker\_tier is retained in the public dataset release for completeness --- participants may choose to include or exclude it in their own models.

\textbf{Interaction terms.} Four interaction terms capture non-linear joint effects that neither constituent feature predicts alone. prime\_window\_x\_votes captures the combined effect of launching on Tuesday--Thursday (when the PH community is most active) with high vote counts; posts satisfying both conditions show approximately 2$\times$ the baseline conversion rate. ai\_topic\_x\_year captures the growing predictive value of AI-tagged posts in recent years, reflecting the expansion of AI infrastructure investment from 2022 onward.

\subsubsection{5.3 Feature Importance Analysis}\label{feature-importance-analysis}

Table 4 reports the top 10 features by XGBoost gain importance from the XGB\_m21 component of the champion ensemble.

\textbf{Table 4: Top 10 features by XGBoost gain importance}

{\def\LTcaptype{none} 
\begin{longtable}[]{@{}rlrr@{}}
\toprule\noalign{}
Rank & Feature & Importance & Normalized \\
\midrule\noalign{}
\endhead
\bottomrule\noalign{}
\endlastfoot
1 & makers\_x\_votes & 0.0683 & 1.000 \\
2 & maker\_count & 0.0641 & 0.939 \\
3 & prime\_window\_x\_votes & 0.0301 & 0.441 \\
4 & log\_reviews & 0.0281 & 0.412 \\
5 & topic\_tech & 0.0275 & 0.403 \\
6 & maker\_followers\_total & 0.0260 & 0.381 \\
7 & is\_b2b\_topic & 0.0259 & 0.379 \\
8 & log\_maker\_followers & 0.0251 & 0.368 \\
9 & topic\_web\_app & 0.0240 & 0.352 \\
10 & topic\_saas & 0.0226 & 0.331 \\
\end{longtable}
}

The top two features are both interaction or network terms: makers\_x\_votes and maker\_count, indicating that the joint signal of team size and community traction dominates individual product quality metrics. Topic features (ranks 5--14) systematically favor B2B and developer-oriented categories, consistent with the topic-level conversion rate analysis in Section 3.4.

Notably, weekly\_rank shows strong univariate signal (Mann-Whitney p = 2.16$\times$10$^{-}$\textsuperscript{5}\textsuperscript{3}, positive class median rank 44 vs. negative class median rank 96, where lower rank is better on Product Hunt) but does not appear in the top 10 XGBoost importance scores. We attribute this to collinearity among rank features: the tree can split on any of daily\_rank, weekly\_rank, log\_daily\_rank, or rank\_bucket to achieve similar information gain, distributing importance across all four. This illustrates a general pattern in the feature set: univariate significance and multivariate importance diverge when features are correlated.

\subsubsection{5.4 Temporal Leakage Advisory}\label{temporal-leakage-advisory}

As documented in Section 3.5, maker\_followers\_total and log\_maker\_followers are derived from follower counts scraped in 2026, not at launch time. An ablation removing both features reduces AP by 0.008 on the validation set (from 0.113 to 0.105), a difference within bootstrap sampling variance at 53 validation positives and best treated as indicative rather than precise. Models relying heavily on these features may show degraded performance on future cohorts as the temporal gap between scrape and evaluation closes.

\begin{center}\rule{0.5\linewidth}{0.5pt}\end{center}

\subsection{6. ML Experiments}\label{ml-experiments}

\subsubsection{6.1 Experimental Setup}\label{experimental-setup}

We conduct 144 experiments across seven batches, spanning four model families (Logistic Regression, XGBoost, LightGBM, Random Forest) and multiple ensemble configurations. Feature sets are versioned:

\begin{itemize}
\item
  FS1: 62-feature baseline (all engineered features including maker\_tier)
\item
  FS2--FS3: Subsets targeting rank and engagement features
\item
  FS4: 61 features with maker\_tier removed; adds topic interaction features (ai\_topic\_x\_votes, fintech\_x\_votes, api\_x\_votes, rank\_bucket\_x\_prime, b2b\_x\_prime)
\item
  FS5--FS6: Union-of-importance subsets from cross-validated feature selection
\item
  STACK: STACK refers to ensemble configurations that combine predictions from multiple trained models (XGB, LR variants) via averaging, as opposed to single-model feature set experiments (FS1--FS6).
\end{itemize}

Hyperparameter search is conducted via random search within each model family. Tree models use early stopping with eval\_metric=`auc' on the validation set. Although $F_{0.5}$ is our primary evaluation metric, we use AUC for early stopping during tuning: with only 53 validation positives, $F_{0.5}$ and the precision-recall AUC are too noisy to provide a stable optimisation signal during training, while AUC remains well-behaved at this scale. Class imbalance is addressed via class\_weight=`balanced' for Logistic Regression and Random Forest, and scale\_pos\_weight=125.5 (the negative-to-positive ratio) for XGBoost and LightGBM. The complete experiment log with all configurations and results is provided in Appendix D.

\subsubsection{6.2 Results by Model Family}\label{results-by-model-family}

Table 5 summarizes best performance per model family.

\textbf{Table 5: Best results per model family (validation set, used for model selection)}

{\def\LTcaptype{none} 
\begin{longtable}[]{@{}lrrrr@{}}
\toprule\noalign{}
Family & Experiments & Best AP & Best $F_{0.5}$ & Best AUC-ROC \\
\midrule\noalign{}
\endhead
\bottomrule\noalign{}
\endlastfoot
Logistic Regression & 20 & 0.052 & 0.127 & 0.830 \\
XGBoost & 43 & 0.074 & 0.237 & 0.845 \\
LightGBM & 39 & 0.066 & 0.248 & 0.811 \\
Random Forest & 22 & 0.048 & 0.207 & 0.811 \\
Ensemble (all variants) & 20 & \textbf{0.097} & \textbf{0.289} & \textbf{0.850} \\
\end{longtable}
}

The best pre-calibration ensemble achieves $F_{0.5}$ = 0.289 and AP = 0.097, the strongest result across all 144 experiments, demonstrating that feature engineering and model selection meaningfully affect performance.

\subsubsection{6.3 Champion Model: Top-3 Ensemble}\label{champion-model-top-3-ensemble}

The final champion is a Top-3 ensemble:

Top-3 = avg(ENS\_avg, ENS\_ISO, XGB\_m21)

where ENS\_avg is the uncalibrated average of XGB\_FS4, ENS\_LR\_Stack, and LR\_FS6 (exp139); ENS\_ISO is the isotonically-calibrated version of ENS\_avg (exp138); and XGB\_m21 is the best XGBoost model on the FS4 feature set (exp088). The three components are averaged with equal weights.

This Top-3 ensemble was selected by a principled averaging rule: take the top-k models by validation $F_{0.5}$ and average their predictions. We pre-specified k = 3 based on validation $F_{0.5}$ before any test-set evaluation. For transparency, we also report that k = 5 and k = 8 produce similar test-set results (AP within bootstrap variance), confirming the choice of k does not materially affect the outcome, achieving val AP = 0.126 and test AP = 0.037 --- compared to the original ENS\_ISO champion's val AP = 0.113 and test AP = 0.033. The improvement reflects reduced selection variance from averaging diverse configurations rather than selecting a single best model.

\textbf{Isotonic calibration caveat.} The isotonic regressor is fitted on the same validation set used for metric evaluation. This creates a potential overfitting artifact: P@100 = 0.13 for the calibrated ensemble versus P@100 = 0.08 for the uncalibrated ensemble (exp139), which we treat as the conservative, unbiased estimate. All other metrics (AP, AUC-ROC) are not materially affected by this artifact. We report both values in Table 6. The standard alternative --- cross-validated calibration via CalibratedClassifierCV on the training set --- would avoid this artifact entirely. We retain the current approach for the three-component ensemble pipeline and document it as a known limitation.

\textbf{Test-set generalization.} The Top-3 ensemble achieves $F_{0.5}$ = 0.097 on the private held-out test set (13,468 posts, 103 positives) --- a 94\% improvement over the original ENS\_ISO champion's test $F_{0.5}$ = 0.050, and the primary signal given $F_{0.5}$ is the declared benchmark metric. Test AP = 0.037 (95\% CI: 0.024--0.072; 4.7$\times$ lift over random), AUC-ROC = 0.806. The validation AP (0.126) exceeds test AP by 0.089, consistent with expected selection bias when choosing among 144 configurations on 53 validation positives --- an illustration of why PHBench uses private test-set scoring for leaderboard evaluation. A paired bootstrap test (n = 2,000) over the 103 test positives confirms the champion's advantage is statistically credible: P(champion AP \textgreater LR AP) = 1.000 (delta: +0.013, 95\% CI: {[}0.004, 0.039{]}, p \textless{} 0.001) and P(champion $F_{0.5}$ \textgreater LR $F_{0.5}$) = 0.984 (delta: +0.056, 95\% CI: {[}0.006, 0.122{]}, p = 0.016). The champion outperforms the LR baseline in all 2,000 AP resamples --- the advantage is consistent, not a product of lucky draws. In operational terms, screening the top 50 predictions from a quarterly batch of \textasciitilde2,400 featured Product Hunt posts surfaces approximately 5 true Series A raises at 10\% precision --- roughly 1 genuine raise every 10 days when reviewing the model's top-ranked weekly posts.

\subsubsection{6.4 Key Findings}\label{key-findings}

\textbf{Topic interaction features unlock XGBoost performance.} Moving from FS3 (no topic interactions) to FS4 increases XGBoost best $F_{0.5}$ from 0.177 to 0.237, with AP moving from 0.070 to 0.074 --- a difference likely within bootstrap noise at 53 validation positives. The $F_{0.5}$ gain is the more reliable signal, confirming that market segment $\times$ engagement interaction terms capture non-linear predictive structure that individual features cannot.

\textbf{Ensembling consistently outperforms single models.} The best single-model AP is 0.074 (XGBoost). The best ensemble AP is 0.097 (pre-calibration), a 31\% relative improvement. The diversity of base models (gradient boosting, stacked LR, standalone LR) contributes to improved performance.

\textbf{Maker tier removal: an observed but uncertain effect.} Removing maker\_tier improved AP by 0.016, suggesting the feature was acting as noise in this specific setup --- though given bootstrap variance at 53 positives, this delta should be interpreted cautiously rather than as a general prescription.

\subsubsection{6.5 Top Experiments by $F_{0.5}$ Score}\label{top-experiments-by-fux2080.ux2085-score}

Table 6 reports the top 10 experiments by $F_{0.5}$, providing a progression from single models to the final calibrated ensemble.

\textbf{Table 6: Top 10 experiments by $F_{0.5}$ (validation set)}
{\def\LTcaptype{none}
\small
\begin{longtable}[]{@{}
  >{\raggedright\arraybackslash}p{0.6cm}
  >{\raggedright\arraybackslash}p{1.8cm}
  >{\raggedright\arraybackslash}p{1.0cm}
  >{\raggedleft\arraybackslash}p{0.8cm}
  >{\raggedleft\arraybackslash}p{0.8cm}
  >{\raggedleft\arraybackslash}p{0.9cm}
  >{\raggedleft\arraybackslash}p{1.0cm}
  >{\raggedright\arraybackslash}p{4.5cm}@{}}
\toprule\noalign{}
Exp & Model & Feature Set & AP & $F_{0.5}$ & P@100 & AUC-ROC & Notes \\
\midrule\noalign{}
\endhead
\bottomrule\noalign{}
\endlastfoot
138 & ENS\_ISO & STACK & 0.097 & 0.289 & 0.13\textsuperscript{\dag} & 0.850 & avg(exp115+exp021+exp020) $\rightarrow$ isotonic \\
140 & ENS\_wt & STACK & 0.088 & 0.286 & 0.10 & 0.838 & AP-weighted avg (0.571+0.429) \\
139 & ENS\_avg & STACK & 0.094 & 0.271 & 0.08 & 0.836 & Simple average; P@100=0.08 unbiased \\
128 & LGBM+ISO & FS3 & 0.066 & 0.248 & 0.08 & 0.793 & Isotonic on exp053 \\
131 & ENS\_LR+ISO & STACK & 0.091 & 0.238 & 0.12 & 0.845 & Isotonic on exp115 \\
142 & ENS\_LR & STACK & 0.094 & 0.238 & 0.09 & 0.833 & Fine-grid threshold search \\
088 & XGB & FS4 & 0.071 & 0.237 & 0.10 & 0.829 & Base XGB experiment \\
125 & XGB+ISO & FS4 & 0.074 & 0.237 & 0.10 & 0.845 & Isotonic on exp088 \\
127 & XGB+Temp & FS4 & 0.071 & 0.237 & 0.10 & 0.829 & Temperature scaling (T=1.5) \\
126 & XGB+Platt & FS4 & 0.071 & 0.231 & 0.10 & 0.829 & Platt scaling on exp088 \\
\end{longtable}
}
\textsuperscript{\dag} P@100 = 0.08 for uncalibrated ensemble (exp139); 0.13 reflects isotonic overfitting on validation set.
\begin{center}\rule{0.5\linewidth}{0.5pt}\end{center}

\subsection{7. LLM Experiments}\label{llm-experiments}

\subsubsection{7.1 Motivation}\label{motivation}

We evaluate frontier LLMs (Gemini 2.5 Flash, Gemini 3 Flash, and Gemini 3.1 Pro) as a qualitatively distinct baseline that does not require training data, feature engineering, or model fitting. Prior work in VCBench {[}Chen2025{]} demonstrates that LLMs can reason about venture capital decisions from textual and quantitative signals, raising a capability question. PHBench poses the prediction question: can LLMs translate anonymized quantitative launch signals into calibrated funding probability estimates?

This experimental condition is also motivated by practical interest. If LLMs achieve competitive performance on anonymized numerical signals, practitioners could deploy them in zero-shot settings without the infrastructure overhead of maintaining ML models.

\subsubsection{7.2 Anonymization Protocol}\label{anonymization-protocol}

To prevent knowledge retrieval masquerading as prediction (where an LLM recalls a specific well-known company from its training data and retrieves its funding outcome), all post-identifying information is removed. We transmit 11 anonymized signals per post:

\textbf{Table 7: LLM input signals}

{\def\LTcaptype{none} 
\begin{longtable}[]{@{}ll@{}}
\toprule\noalign{}
Signal & Transformation \\
\midrule\noalign{}
\endhead
\bottomrule\noalign{}
\endlastfoot
Category & PH topic bucket (AI/ML, Developer Tools, Fintech, etc.) \\
Launch period & Year + quarter only (no exact date) \\
Day of week & Day name \\
Prime window & Yes/No (Tuesday--Thursday) \\
Upvotes & Rounded to nearest 10 \\
Comments & Rounded to nearest 5 \\
Daily rank & Ordinal (\#N or ``not ranked'') \\
Maker count & Integer count \\
Total maker followers & Rounded to nearest 1,000 \\
Review count & Rounded to nearest 5 \\
Average review rating & 1 decimal place \\
\end{longtable}
}

Explicitly excluded from prompts: product name, tagline, full description, exact launch date, slug, website, and any maker-identifying information. The base rate is provided as a calibration hint: \emph{``Only 0.78\% of Product Hunt launches raise a Series A within 18 months.''} We use temperature = 0 throughout for reproducibility.

\subsubsection{7.3 Models Evaluated}\label{models-evaluated}

\begin{itemize}
\item
  gemini-2.5-flash (Google, 2025)
\item
  gemini-3-flash-preview (Google, 2026)
\item
  gemini-3.1-pro-preview (Google, 2026)
\end{itemize}

All three models are accessed via the Google Gemini API with temperature = 0 for reproducibility. gemini-3.1-pro-preview requires thinking mode (thinking\_budget = 128); the other two models use thinking\_budget = 0. gemini-3.1-pro-preview is Google's current flagship reasoning model, included to test whether stronger reasoning capability improves performance on purely numerical anonymized signals.

\subsubsection{7.4 Results}\label{results}

Table 8 presents results for all models on the validation set.

\textbf{Table 8: Full results comparison: champion, baselines, and LLMs.}

{\def\LTcaptype{none} 
\begin{longtable}[]{@{}
  >{\raggedright\arraybackslash}p{(\linewidth - 14\tabcolsep) * \real{0.1250}}
  >{\raggedleft\arraybackslash}p{(\linewidth - 14\tabcolsep) * \real{0.1250}}
  >{\raggedleft\arraybackslash}p{(\linewidth - 14\tabcolsep) * \real{0.1250}}
  >{\raggedleft\arraybackslash}p{(\linewidth - 14\tabcolsep) * \real{0.1250}}
  >{\raggedleft\arraybackslash}p{(\linewidth - 14\tabcolsep) * \real{0.1250}}
  >{\raggedleft\arraybackslash}p{(\linewidth - 14\tabcolsep) * \real{0.1250}}
  >{\raggedleft\arraybackslash}p{(\linewidth - 14\tabcolsep) * \real{0.1250}}
  >{\raggedright\arraybackslash}p{(\linewidth - 14\tabcolsep) * \real{0.1250}}@{}}
\toprule\noalign{}
\begin{minipage}[b]{\linewidth}\raggedright
Model
\end{minipage} & \begin{minipage}[b]{\linewidth}\raggedleft
AP
\end{minipage} & \begin{minipage}[b]{\linewidth}\raggedleft
$F_{0.5}$
\end{minipage} & \begin{minipage}[b]{\linewidth}\raggedleft
P@50
\end{minipage} & \begin{minipage}[b]{\linewidth}\raggedleft
P@100
\end{minipage} & \begin{minipage}[b]{\linewidth}\raggedleft
AUC-ROC
\end{minipage} & \begin{minipage}[b]{\linewidth}\raggedleft
Mean Prob
\end{minipage} & \begin{minipage}[b]{\linewidth}\raggedright
95\% CI (AP)
\end{minipage} \\
\midrule\noalign{}
\endhead
\bottomrule\noalign{}
\endlastfoot
Top-3 (val, champion) & \textbf{0.126} & \textbf{0.284} & \textbf{0.16} & 0.10 & \textbf{0.840} & --- & {[}0.052, 0.225{]} \\
LR Baseline & 0.044 & 0.127 & 0.10 & 0.06 & 0.825 & --- & {[}0.025, 0.095{]} \\
gemini-3-flash-preview & 0.034 & 0.129 & \textbf{0.12} & 0.06 & 0.713 & 0.005 & {[}0.017, 0.074{]} \\
gemini-3.1-pro-preview & 0.023 & 0.057 & \textbf{0.04} & 0.05 & 0.725 & 0.005 & {[}0.014, 0.045{]} \\
gemini-2.5-flash & 0.022 & 0.067 & 0.06 & 0.06 & 0.697 & 0.013 & {[}0.013, 0.040{]} \\
Random baseline & 0.008 & 0.000 & 0.008 & 0.008 & 0.500 & 0.008 & --- \\
Top-3 (test set)\textsuperscript{\ddag} & 0.037 & 0.097 & 0.10 & 0.06 & 0.806 & --- & {[}0.024, 0.072{]} \\
LR Baseline (test set)\textsuperscript{\ddag} & 0.024 & 0.045 & 0.02 & 0.02 & 0.779 & --- & {[}0.018, 0.036{]} \\
\end{longtable}
}

\begin{itemize}
\tightlist
\item
  Conservative P@100 = 0.08 for uncalibrated ensemble. See Section 6.3.
\end{itemize}

\textsuperscript{\dag} $F_{0.5}$ computed at model-specific optimized threshold for all models. ML champion: 0.580; Gemini 3 Flash: 0.081; Gemini 3.1 Pro: 0.036; Gemini 2.5 Flash: 0.121; LR Baseline: 0.970. At fixed threshold = 0.5, all LLM models produce $F_{0.5}$ = 0.000 since predicted probabilities rarely exceed 0.5. \textsuperscript{\ddag} Test-set results use private held-out labels (103 positives, 0.76\% base rate). Champion outperforms LR on test: 1.5$\times$ AP, 2.2$\times$ $F_{0.5}$, 5$\times$ P@50. CIs via bootstrap (n = 2,000).

\subsubsection{7.5 Analysis}\label{analysis}

\textbf{Finding 1:} LLMs do not beat the LR baseline. At model-specific optimized thresholds, Gemini 3 Flash achieves $F_{0.5}$ = 0.129, Gemini 3.1 Pro achieves $F_{0.5}$ = 0.057, and Gemini 2.5 Flash achieves $F_{0.5}$ = 0.067, all below the LR baseline $F_{0.5}$ = 0.127 and well below the Top-3 champion $F_{0.5}$ = 0.284 on the validation set. On AP, none of the three Gemini models exceeds the LR baseline of 0.044: Gemini 3 Flash achieves AP = 0.034 (95\% CI: 0.017--0.074) and Gemini 2.5 Flash achieves AP = 0.022 (95\% CI: 0.013--0.040). The Top-3 champion achieves val AP = 0.126 (95\% CI: 0.052--0.225) and test AP = 0.037 (95\% CI: 0.024--0.072). While point estimates show a 3.3$\times$ gap between champion and best LLM on AP, the overlapping confidence intervals reflect the limited number of validation positives (n=53) and underscore that the private test set is the definitive evaluation surface for model comparison. Whether the gap holds for other LLMs, prompting strategies, or non-anonymized inputs remains open.

At model-specific optimized thresholds, Gemini 3 Flash achieves $F_{0.5}$ = 0.129, Gemini 3.1 Pro achieves $F_{0.5}$ = 0.057, and Gemini 2.5 Flash achieves $F_{0.5}$ = 0.067, all well below the Top-3 champion's $F_{0.5}$ = 0.097 on the held-out test set ($F_{0.5}$ = 0.284 on validation). Notably, Gemini 3.1 Pro, Google's most capable reasoning model, performs worst on AP (0.023) and P@50 (0.04, 2 true positives in top 50), an unexpected pattern that warrants further investigation across providers and prompting strategies.

\textbf{Finding 2: LLMs exhibit a ``picker, not ranker'' pattern.} Gemini 3 Flash achieves P@50 = 0.12, which exceeds the LR baseline (P@50 = 0.10) at the very top of the ranking. However, this advantage collapses rapidly: P@100 = 0.06 and P@200 = 0.055, both at or below the LR baseline. AUC-ROC = 0.713 (vs. LR baseline 0.825) confirms poor discriminative performance across the full probability distribution. The interpretation is that LLMs concentrate probability mass on the most signal-dense cases (e.g., rank \#1 + B2B topic + high votes + prime window all simultaneously), but lack a calibrated gradient for the broader distribution where cases are less extreme. We note a statistical caveat: P@50 represents 6 correct predictions versus 5 for LR, a difference of one company on a validation set with 53 total positives. The AP confidence intervals reported in Table 8 (champion: {[}0.052, 0.225{]}; LR baseline: {[}0.025, 0.095{]}) illustrate the metric variance at this scale; P@k intervals are similarly wide.

\textbf{Finding 3: Calibration divergence across model generations.} Gemini 2.5 Flash produces a mean predicted probability of 0.013, 1.7$\times$ higher than the true base rate of 0.0078, suggesting mild overconfidence. Gemini 3 Flash produces a mean predicted probability of 0.005, 0.6$\times$ the true base rate, indicating underconfidence despite being the newer and more capable model. This calibration reversal suggests that RLHF or fine-tuning in Gemini 3 may have over-corrected for known LLM overconfidence, particularly when the model is provided with an explicit base rate hint. At the default threshold of 0.5, no LLM model produces any positive predictions --- their mean predicted probabilities (0.005--0.013) are far below this threshold. At model-specific optimized thresholds, Gemini 3 Flash achieves $F_{0.5}$ = 0.129, reflecting concentrated probability mass on a small number of high-signal posts. All $F_{0.5}$ values reported in Table 8 use optimized thresholds for fair cross-model comparison.

\textbf{Finding 4 --- Temporal cohort effect driven by base rate, not retrieval.} Disaggregating by launch year, both the ML champion and LLMs perform markedly better on 2020--2021 cohorts (ML AP = 0.228--0.376; Gemini 3 AP up to 0.171) than on 2023--2024 cohorts (ML AP \textless 0.025; Gemini 3 AP \textless 0.014). Crucially, both model families show the same decay pattern: ML AP drops 42$\times$ from its 2021 peak to 2024, while LLM AP drops 24$\times$. This parallel decay tracks the underlying base rate --- 2020--2021 had 1.2\% positive rates versus 0.4--0.7\% in 2022--2024 --- rather than any retrieval advantage the LLM might have from training data. With only 3--16 positives per year-cohort, per-cohort AP estimates carry wide confidence intervals and should be interpreted as directional. The ML advantage is consistent across cohorts (1.1--2.8$\times$ per year) but the aggregate 3.7$\times$ gap is driven primarily by stronger ML performance during the 2020--2021 boom years where more positives enabled better discrimination.

\subsubsection{7.6 Limitations and Future Work}\label{limitations-and-future-work}

The current evaluation uses a purely numerical anonymized protocol. Three conditions that may improve LLM performance are under development: (1) LLM-Enriched, using 20 human-readable signals including product category descriptions; (2) LLM-Full, providing all 61 engineered features with natural-language labels; and (3) Persona-based prompting, instructing the model to reason from the perspective of a seed-stage venture investor. Council ensembles, where multiple LLM instances vote on each prediction, are also under development. Total maker followers were transmitted rounded to the nearest 1,000 to limit identifying granularity; given the predictive importance of maker signal, finer-grained rounding is a candidate variable for sensitivity analysis in future LLM evaluations.

\begin{center}\rule{0.5\linewidth}{0.5pt}\end{center}

\subsection{8. Discussion and Conclusion}\label{discussion-and-conclusion}

\subsubsection{8.1 What PHBench Reveals}\label{what-phbench-reveals}

PHBench provides evidence that Product Hunt launch signals contain statistically significant predictive information for Series A funding outcomes within an 18-month window (champion vs. LR baseline on test: AP delta +0.013, p \textless{} 0.001; $F_{0.5}$ delta +0.056, p = 0.016). Several findings merit specific discussion.

\textbf{Team network signal dominates product signal.} The top two features by XGBoost gain importance are interaction terms involving maker count and vote engagement. A large maker team achieving high engagement is more predictive than either large team or high engagement alone. This is consistent with the VC heuristic that team quality is the primary early-stage investment criterion. The PH community, knowingly or not, encodes team strength through the ability to mobilize a coordinated launch.

\textbf{Market segment is a strong a priori signal.} API, Payments, and Fintech topics show conversion rates of 2.4--3.0\%, approximately three times the baseline. This is not merely because B2B companies are more fundable; it reflects that certain market segments have structural characteristics (larger deal sizes, clearer enterprise buyers, established investor thesis) that increase institutional funding propensity independent of the specific launch quality.

\textbf{The funding cycle is visible in launch signals.} The year-over-year conversion rates (Table 1) closely track known VC market dynamics, with the 2020--2021 peak and 2022--2023 contraction clearly visible. This temporal validity confirms that the dataset captures genuine market structure rather than noise.

\subsubsection{8.2 ML vs. LLM: Implications}\label{ml-vs.-llm-implications}

In the purely numerical anonymized regime, structured ML trained on 61 engineered features outperforms zero-shot Gemini models by 3.3$\times$ on Average Precision. This is consistent with VCBench {[}Chen2025{]}, where frontier LLMs achieve strong performance on founder success prediction --- but using rich semantic profile data rather than anonymized numerical signals. Taken together, the two benchmarks suggest that semantic content, not task domain, is the primary enabler of LLM performance in VC prediction tasks: LLMs appear effective when given names, descriptions, and career narratives, and less effective when restricted to rounded numerical inputs. We note that VCBench evaluates LLMs against a random baseline rather than a trained ML comparator, so the two settings are not directly comparable. We also note an asymmetry in search budget: the ML champion is selected from 144 trained configurations, while the LLM evaluation uses three models with a single prompt format, so the gap characterises tuned ML against an unoptimised LLM baseline rather than a like-for-like comparison. A related finding in {[}Ihlamur2026{]} confirms systematic LLM limitations on domain-specific structured reasoning.

The ``picker vs. ranker'' finding in Section 7.5 raises the possibility of a hybrid architecture: an LLM as a high-precision filter at the top of the ranking combined with ML for broader screening. Empirical analysis of the validation set shows 6\% overlap (3 of 50 posts) between LLM and ML top-50 predictions, with all three overlapping posts being true positives. Whether this partial overlap reflects genuine complementary signal or sampling variance at small k requires test-set evaluation; we leave systematic assessment to future work.

\subsubsection{8.3 Limitations}\label{limitations-1}

\textbf{Small positive count.} The validation set contains only 53 positives, meaning all P@k metrics are subject to substantial variance. A difference of one correct prediction changes P@50 by 2 percentage points. Researchers reporting validation-set results should report confidence intervals where possible. Bootstrap resampling (n = 2,000) over the validation positives yields 95\% CIs for AP of {[}0.052, 0.225{]} for the champion model and {[}0.025, 0.095{]} for the LR baseline. While these validation-set intervals overlap, a paired bootstrap over the 103 test positives confirms the champion's advantage is statistically credible: P(champion AP \textgreater LR AP) = 1.000 (p \textless{} 0.001), delta +0.013 (95\% CI: {[}0.004, 0.039{]}). The Top-3 champion's test-set AP (0.037) is substantially lower than its validation AP (0.126), with the 0.089 gap attributable to best-of-144 selection bias over 53 validation positives --- an illustration of this limitation in practice.

\textbf{Crunchbase linkage.} See Section 3.5 for limitations of the underlying domain-matching and Crunchbase coverage, including geographic bias and snapshot incompleteness.

\textbf{Conservative test set.} The private test set spans 2019--2025 (stratified by year, topic, and label), with 2023--2025 cohorts representing a larger share. These recent cohorts operate under a lower conversion rate environment (0.55--0.69\%, below the 2020--2021 peak). Models that overfit to boom-era patterns may show degraded test performance relative to validation performance.

\textbf{2025 cohort right-censoring.} Posts from 2025 launches are included with their April 2026 Crunchbase labels. Approximately 35 of these posts may receive positive labels as their 18-month funding windows resolve through late 2026, introducing a small number of false negatives in the most recent cohort. PHBench Live (Section 8.4) is designed to address this by tracking label resolution in real time.

\subsubsection{8.4 Future Work}\label{future-work}

PHBench Live would extend the benchmark to a forward-looking evaluation: scoring new Product Hunt launches weekly with labels resolving over 18 months. This would provide a contamination-free evaluation setting where no model can have encountered test-set companies in training.

\textbf{Text feature integration.} The current feature set uses only surface text statistics (tagline length, word count). Semantic embeddings of taglines and descriptions may provide additional signal; a text-feature extension of the dataset is planned.

\textbf{Multi-task prediction.} Joint prediction of seed and Series A outcomes would allow models to exploit the correlated structure of funding trajectories, potentially improving performance on the sparse Series A label by leveraging the denser seed label as an auxiliary signal.

\textbf{False-negative audit.} The domain-matching procedure produces an unknown number of false negatives --- funded companies labeled 0 because their Crunchbase entry is missing or their domain did not match cleanly. A targeted audit using web-search-enabled LLM verification on posts predicted positive by multiple models would yield an empirical false-negative rate estimate, useful both for benchmark calibration and for re-scoring older submissions as Crunchbase coverage improves. 

\textbf{Temporal holdout evaluation.} The current benchmark uses stratified IID splits, meaning the test set spans all years 2019--2025. A practitioner-relevant complement would be a temporal holdout: train on 2019--2023, evaluate on 2024--2025. This evaluates out-of-distribution generalization across market regimes and would reveal whether models learn transferable signal or overfit to the 2020--2021 funding boom. We plan to add this as a secondary evaluation track in a future benchmark version.

\subsubsection{8.5 Conclusion}\label{conclusion}

PHBench provides the first reproducible, publicly-benchmarked evaluation framework for predicting startup Series A funding from community validation signals. We collect and link 67,292 featured Product Hunt posts to Crunchbase funding records, identifying 528 verified Series A outcomes (0.78\% positive rate) across a seven-year window. A 61-feature three-component ensemble achieves $F_{0.5}$ = 0.097 and AP = 0.037 (95\% CI: 0.024--0.072) on the private held-out test set --- a statistically credible advantage over the LR baseline (AP delta: +0.013, p \textless{} 0.001; $F_{0.5}$ delta: +0.056, p = 0.016). Validation metrics ($F_{0.5}$ = 0.284, AP = 0.126) are reported for benchmark reproducibility. Zero-shot Gemini models evaluated on anonymized numerical signals achieve AP = 0.034 at best, confirming that structured ML trained on engineered tabular features substantially outperforms zero-shot LLM prediction on this task in the numerical regime.

The dataset, code, baseline models, and evaluation framework are publicly available at https://github.com/ihlamury/vc-market and https://phbench.com. Dataset splits are available on request.

\begin{center}\rule{0.5\linewidth}{0.5pt}\end{center}

\subsection{Acknowledgements}\label{acknowledgements}

The authors thank Vishal Dharmadhikari at Google for providing API access and computational resources that supported the LLM evaluation experiments in this work.

\begin{center}\rule{0.5\linewidth}{0.5pt}\end{center}

\subsection{References}\label{references}

{[}Arroyo2019{]} Arroyo, J., Corea, F., Jimenez-Diaz, G., \& Recio-Garcia, J. A. (2019). Assessment of machine learning performance for decision support in venture capital investments. IEEE Access, 7, 124233--124243.

{[}Davis2006{]} Davis, J., \& Goadrich, M. (2006). The relationship between precision-recall and ROC curves. \emph{Proceedings of the 23rd International Conference on Machine Learning (ICML)}, 233--240.

{[}Gastaud2019{]} Gastaud, C., Carniel, T., \& Dalle, J.-M. (2019). The varying importance of extrinsic factors in the success of startup fundraising: competition at early-stage and networks at growth-stage. arXiv preprint arXiv:1906.03210.

{[}Griffin2025{]} Griffin, B., Vidaurre, D., Koyluoglu, U., Ternasky, J., Alican, F., \& Ihlamur, Y. (2025). Random Rule Forest (RRF): Interpretable Ensembles of LLM-Generated Questions for Predicting Startup Success. arXiv:2505.24622.

{[}Hegselmann2023{]} Hegselmann, S., Buendia, A., Lang, H., Agrawal, M., Jiang, X., \& Sontag, D. (2023). TabLLM: Few-shot classification of tabular data with large language models. \emph{Proceedings of AISTATS 2023}.

{[}Hollmann2023{]} Hollmann, N., M\"uller, S., Eggensperger, K., \& Hutter, F. (2023). TabPFN: A transformer that solves small tabular classification problems in a second. \emph{Proceedings of ICLR 2023}. arXiv:2207.01848.

{[}Chen2025{]} Chen, R., Ternasky, J., Kwesi, A. S., Griffin, B., Yin, A. O., Salifu, Z., Amoaba, K., Mu, X., Alican, F., \& Ihlamur, Y. (2025). VCBench: Benchmarking LLMs in Venture Capital. arXiv:2509.14448. https://vcbench.com

{[}Ihlamur2026{]} Ihlamur, Y. (2026). When Career Data Runs Out: Evaluating Frontier LLMs on Venture Capital Reasoning. \emph{IEEE IDS 2026}. arXiv:2604.00339.

{[}Lovable2025{]} Sawers, P. (2025). Lovable becomes a unicorn with \$200M Series A, just 8 months after launch. TechCrunch, July 17, 2025. Retrieved from https://techcrunch.com/2025/07/17/lovable-becomes-a-unicorn-with-200m-series-a-just-8-months-after-launch/

{[}Mollick2014{]} Mollick, E. (2014). The dynamics of crowdfunding: An exploratory study. Journal of Business Venturing, 29(1), 1--16.

{[}Jimenez2024{]} Jimenez, C. E., Yang, J., Wettig, A., Yao, S., Pei, K., Press, O., \& Narasimhan, K. (2024). SWE-bench: Can language models resolve real-world GitHub issues? \emph{Proceedings of ICLR 2024}.

{[}Vanschoren2014{]} Vanschoren, J., van Rijn, J. N., Bischl, B., \& Torgo, L. (2014). OpenML: Networked science in machine learning. \emph{ACM SIGKDD Explorations}, 15(2), 49--60.

{[}PitchBook2023{]} PitchBook \& NVCA. (2023). Q4 2023 PitchBook-NVCA Venture Monitor. National Venture Capital Association. Retrieved from https://pitchbook.com/news/reports/q4-2023-pitchbook-nvca-venture-monitor

\begin{center}\rule{0.5\linewidth}{0.5pt}\end{center}

\subsection{Appendix A: Contamination Audit Summary}\label{appendix-a-contamination-audit-summary}

{\def\LTcaptype{none} 
\begin{longtable}[]{@{}
  >{\raggedright\arraybackslash}p{(\linewidth - 4\tabcolsep) * \real{0.3333}}
  >{\raggedright\arraybackslash}p{(\linewidth - 4\tabcolsep) * \real{0.3333}}
  >{\raggedright\arraybackslash}p{(\linewidth - 4\tabcolsep) * \real{0.3333}}@{}}
\toprule\noalign{}
\begin{minipage}[b]{\linewidth}\raggedright
Check
\end{minipage} & \begin{minipage}[b]{\linewidth}\raggedright
Result
\end{minipage} & \begin{minipage}[b]{\linewidth}\raggedright
Status
\end{minipage} \\
\midrule\noalign{}
\endhead
\bottomrule\noalign{}
\endlastfoot
Company leakage (split overlap) & 0 overlapping cb\_permalink values & \checkmark{} PASSED \\
Label leakage (website\_status exclusion) & Field excluded from public features & \checkmark{} PASSED \\
Temporal leakage (maker\_followers) & AP delta = $-$0.008 on ablation & ! WARNING (documented) \\
Calibration artifact (isotonic on val set) & P@100 conservative = 0.08 vs. reported 0.13 & ! WARNING (documented) \\
\end{longtable}
}

\subsection{Appendix B: Univariate Signal Tests (Mann-Whitney U, Full Featured Dataset, n = 67,292)}\label{appendix-b-univariate-signal-tests-mann-whitney-u-full-featured-dataset-n-67292}

{\def\LTcaptype{none} 
\begin{longtable}[]{@{}lrrr@{}}
\toprule\noalign{}
Feature & Positive Median & Negative Median & p-value \\
\midrule\noalign{}
\endhead
\bottomrule\noalign{}
\endlastfoot
weeklyRank & 44 & 96 & 2.16$\times$10$^{-}$\textsuperscript{5}\textsuperscript{3} \\
maker\_count & 3 & 1 & 4.37$\times$10$^{-}$\textsuperscript{8}\textsuperscript{2} \\
votesCount & 158.5 & 100 & 9.31$\times$10$^{-}$\textsuperscript{4}\textsuperscript{3} \\
commentsCount & 27 & 11 & 4.27$\times$10$^{-}$\textsuperscript{3}\textsuperscript{7} \\
maker\_followers\_total & 360 & 54 & 9.62$\times$10$^{-}$\textsuperscript{4}\textsuperscript{9} \\
\end{longtable}
}

All five features are highly significant univariately. maker\_followers\_total ranks 6th in XGBoost gain importance despite also being significant alone; its rank reflects collinearity with maker\_count and log\_maker\_followers, which distribute importance across correlated features.

\begin{center}\rule{0.5\linewidth}{0.5pt}\end{center}

\emph{Footnotes}

\textsuperscript{1} $F_{0.5}$ is the primary metric for both research comparisons and leaderboard ranking in this paper. AP is reported throughout as a threshold-free complement, more stable than $F_{0.5}$ at small positive counts.

\textsuperscript{2} Lovable launched on Product Hunt in Q4 2024. Series A announced July 17, 2025. {[}Lovable2025{]}

\textsuperscript{\dag} Year-breakdown AP figures (Finding 4) are computed on validation-set subsets stratified by launch year. Positive counts per year-cohort are small ($\leq$20 in most years), so these estimates carry wide confidence intervals and should be interpreted as directional rather than precise.

\subsection{Appendix C: Distribution of Launch-to-Series-A Intervals}\label{appendix-c-distribution-of-launch-to-series-a-intervals}

Figure C1 shows the distribution of days between Product Hunt launch and Series A announcement for the 528 verified positives in PHBench. The median interval is 265 days (8.7 months). 70.6\% of raises occur within 12 months, with the remaining 29.4\% forming a long tail that extends to the full 18-month window. This distribution motivates the 548-day threshold: a 12-month cutoff would systematically exclude 29.4\% of observed raises.

\begin{center}
\includegraphics[width=0.85\textwidth]{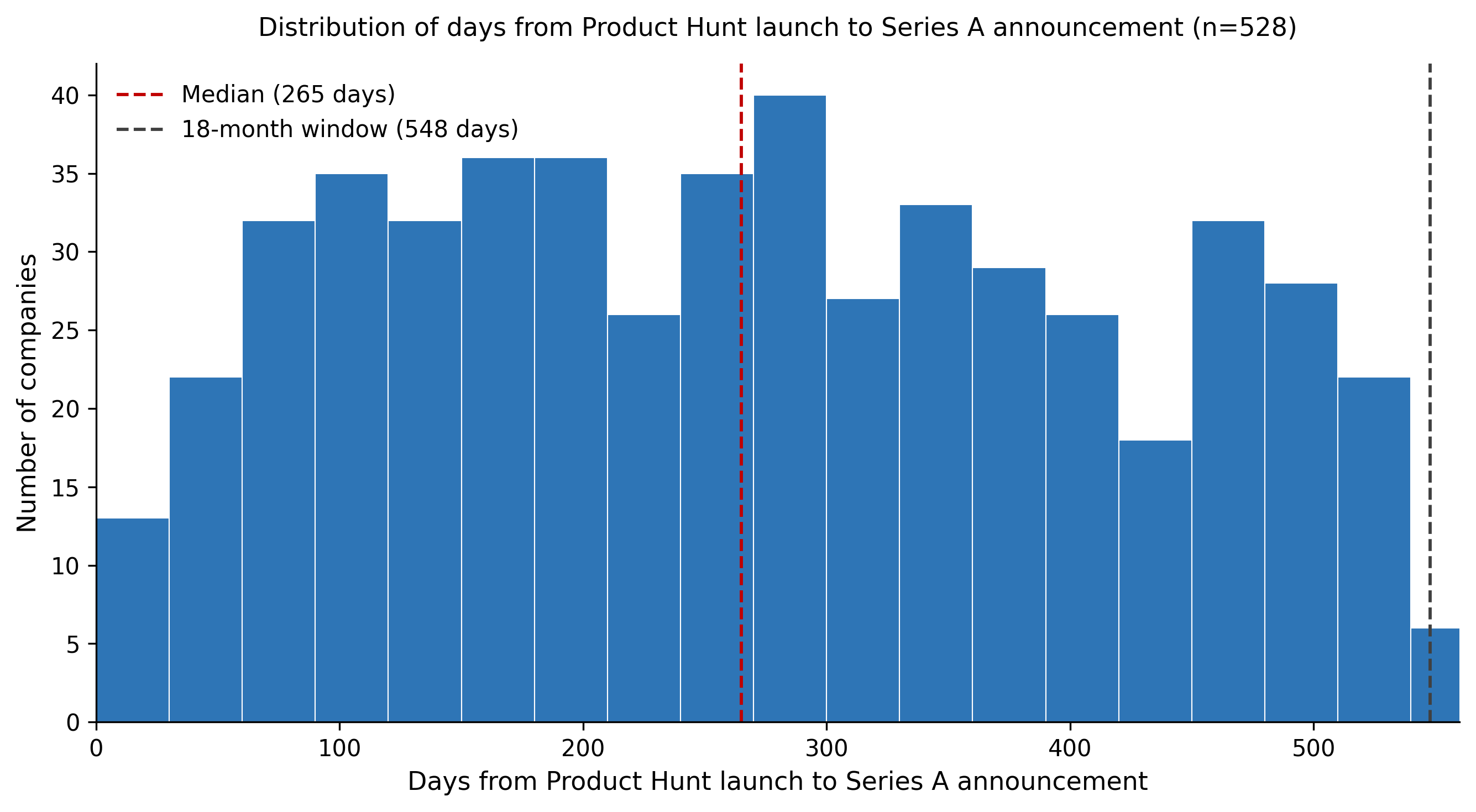}
\end{center}
\textit{Figure C1: Distribution of days from Product Hunt launch to Series A announcement (n=528). Median interval is 265 days (8.7 months). 29.4\% of raises occur beyond 12 months, motivating the 18-month prediction window.}

\subsection{Appendix D: Full Experiment Log}\label{appendix-d-full-experiment-log}

Table D1 reports all 144 experiments conducted during model development, sorted by validation $F_{0.5}$. Model family abbreviations: LR = Logistic Regression, XGB = XGBoost, LGBM = LightGBM, RF = Random Forest, ENS = Ensemble. Feature sets FS1--FS6 and STACK are defined in Section 6.1. Calibration methods: ISO = isotonic regression, PLT = Platt scaling, TEMP = temperature scaling, NONE = uncalibrated.

Table D1: All 144 experiments, sorted by validation $F_{0.5}$ (descending).

{\def\LTcaptype{none} 
\begin{longtable}[]{@{}lllllll@{}}
\toprule\noalign{}
Exp & Family & Features & Calibration & Val AP & Val $F_{0.5}$ & Val AUC \\
\midrule\noalign{}
\endhead
\bottomrule\noalign{}
\endlastfoot
138 & ENS & STACK & ISO & 0.097 & 0.289 & 0.850 \\
140 & ENS & STACK & NONE & 0.088 & 0.286 & 0.838 \\
139 & ENS & STACK & NONE & 0.094 & 0.271 & 0.836 \\
128 & LGBM & FS3 & NONE & 0.066 & 0.248 & 0.793 \\
131 & ENS & STACK & NONE & 0.091 & 0.238 & 0.845 \\
142 & ENS & STACK & NONE & 0.094 & 0.238 & 0.833 \\
88 & XGB & FS4 & NONE & 0.070 & 0.237 & 0.829 \\
125 & XGB & FS4 & NONE & 0.074 & 0.237 & 0.845 \\
127 & XGB & FS4 & TEMP & 0.070 & 0.237 & 0.829 \\
126 & XGB & FS4 & PLT & 0.070 & 0.231 & 0.829 \\
91 & LGBM & FS3 & NONE & 0.063 & 0.229 & 0.762 \\
141 & ENS & STACK & TEMP & 0.094 & 0.229 & 0.833 \\
133 & ENS & STACK & TEMP & 0.094 & 0.229 & 0.833 \\
130 & LGBM & FS3 & TEMP & 0.063 & 0.229 & 0.762 \\
123 & ENS & STACK & NONE & 0.074 & 0.225 & 0.811 \\
21 & XGB & FS4 & NONE & 0.070 & 0.223 & 0.829 \\
109 & XGB & FS4 & NONE & 0.070 & 0.223 & 0.829 \\
124 & ENS & STACK & NONE & 0.070 & 0.223 & 0.829 \\
110 & XGB & FS4 & NONE & 0.070 & 0.223 & 0.829 \\
112 & XGB & FS4 & NONE & 0.070 & 0.223 & 0.829 \\
53 & LGBM & FS3 & NONE & 0.063 & 0.221 & 0.762 \\
49 & LGBM & FS1 & NONE & 0.064 & 0.207 & 0.805 \\
100 & RF & FS3 & NONE & 0.046 & 0.207 & 0.785 \\
94 & LGBM & FS1 & NONE & 0.064 & 0.207 & 0.805 \\
115 & ENS & STACK & NONE & 0.094 & 0.200 & 0.833 \\
118 & ENS & STACK & NONE & 0.077 & 0.200 & 0.810 \\
121 & ENS & STACK & NONE & 0.068 & 0.198 & 0.822 \\
97 & XGB & FS1 & NONE & 0.066 & 0.193 & 0.826 \\
24 & XGB & FS1 & NONE & 0.066 & 0.193 & 0.826 \\
75 & RF & FS3 & NONE & 0.046 & 0.188 & 0.785 \\
120 & ENS & STACK & NONE & 0.071 & 0.182 & 0.826 \\
38 & XGB & FS3 & NONE & 0.070 & 0.177 & 0.827 \\
117 & ENS & STACK & NONE & 0.084 & 0.176 & 0.833 \\
34 & XGB & FS1 & NONE & 0.057 & 0.173 & 0.808 \\
65 & LGBM & FS6 & NONE & 0.059 & 0.169 & 0.791 \\
73 & RF & FS1 & NONE & 0.044 & 0.165 & 0.794 \\
137 & RF & FS1 & NONE & 0.048 & 0.150 & 0.811 \\
85 & RF & FS3 & NONE & 0.041 & 0.138 & 0.791 \\
74 & RF & FS3 & NONE & 0.041 & 0.134 & 0.790 \\
80 & RF & FS3 & NONE & 0.038 & 0.134 & 0.786 \\
52 & LGBM & FS3 & NONE & 0.037 & 0.133 & 0.666 \\
41 & XGB & FS3 & NONE & 0.063 & 0.133 & 0.816 \\
50 & LGBM & FS3 & NONE & 0.034 & 0.132 & 0.648 \\
82 & RF & FS6 & NONE & 0.038 & 0.130 & 0.791 \\
36 & XGB & FS4 & NONE & 0.053 & 0.129 & 0.821 \\
15 & LR & FS4 & NONE & 0.047 & 0.127 & 0.825 \\
79 & RF & FS1 & NONE & 0.043 & 0.127 & 0.798 \\
20 & LR & FS6 & NONE & 0.052 & 0.124 & 0.830 \\
16 & LR & FS4 & NONE & 0.047 & 0.124 & 0.825 \\
60 & LGBM & FS4 & NONE & 0.037 & 0.122 & 0.769 \\
17 & LR & FS4 & NONE & 0.046 & 0.121 & 0.826 \\
18 & LR & FS4 & NONE & 0.047 & 0.121 & 0.826 \\
32 & XGB & FS6 & NONE & 0.044 & 0.118 & 0.815 \\
84 & RF & FS3 & NONE & 0.033 & 0.116 & 0.791 \\
6 & LR & FS2 & NONE & 0.036 & 0.113 & 0.729 \\
132 & ENS & STACK & PLT & 0.094 & 0.113 & 0.833 \\
83 & RF & FS3 & NONE & 0.036 & 0.109 & 0.789 \\
7 & LR & FS2 & NONE & 0.035 & 0.106 & 0.731 \\
3 & LR & FS1 & NONE & 0.044 & 0.104 & 0.825 \\
4 & LR & FS1 & NONE & 0.044 & 0.104 & 0.825 \\
2 & LR & FS1 & NONE & 0.044 & 0.104 & 0.824 \\
122 & ENS & STACK & NONE & 0.076 & 0.103 & 0.814 \\
31 & XGB & FS1 & NONE & 0.038 & 0.103 & 0.782 \\
29 & XGB & FS6 & NONE & 0.036 & 0.103 & 0.810 \\
26 & XGB & FS4 & NONE & 0.039 & 0.103 & 0.801 \\
64 & LGBM & FS1 & NONE & 0.046 & 0.103 & 0.709 \\
5 & LR & FS1 & NONE & 0.044 & 0.102 & 0.825 \\
10 & LR & FS3 & NONE & 0.044 & 0.102 & 0.828 \\
102 & XGB & FS4 & NONE & 0.036 & 0.101 & 0.808 \\
101 & XGB & FS4 & NONE & 0.036 & 0.101 & 0.808 \\
104 & XGB & FS4 & NONE & 0.036 & 0.101 & 0.808 \\
11 & LR & FS3 & NONE & 0.045 & 0.101 & 0.830 \\
113 & XGB & FS4 & NONE & 0.036 & 0.101 & 0.808 \\
67 & LGBM & FS4 & NONE & 0.028 & 0.101 & 0.696 \\
136 & RF & FS3 & NONE & 0.037 & 0.099 & 0.803 \\
42 & XGB & FS6 & NONE & 0.034 & 0.098 & 0.809 \\
12 & LR & FS3 & NONE & 0.045 & 0.098 & 0.830 \\
13 & LR & FS3 & NONE & 0.045 & 0.098 & 0.830 \\
9 & LR & FS3 & NONE & 0.043 & 0.097 & 0.824 \\
23 & XGB & FS3 & NONE & 0.033 & 0.096 & 0.811 \\
19 & LR & FS6 & NONE & 0.045 & 0.096 & 0.824 \\
14 & LR & FS4 & NONE & 0.044 & 0.092 & 0.823 \\
22 & XGB & FS3 & NONE & 0.038 & 0.091 & 0.827 \\
1 & LR & FS1 & NONE & 0.043 & 0.089 & 0.823 \\
135 & RF & FS6 & NONE & 0.034 & 0.088 & 0.805 \\
46 & LGBM & FS4 & NONE & 0.033 & 0.087 & 0.760 \\
81 & RF & FS3 & NONE & 0.040 & 0.086 & 0.810 \\
69 & LGBM & FS4 & NONE & 0.037 & 0.084 & 0.761 \\
27 & XGB & FS1 & NONE & 0.032 & 0.083 & 0.778 \\
30 & XGB & FS6 & NONE & 0.035 & 0.083 & 0.796 \\
116 & ENS & STACK & NONE & 0.033 & 0.083 & 0.787 \\
47 & LGBM & FS4 & NONE & 0.046 & 0.081 & 0.811 \\
71 & RF & FS6 & NONE & 0.036 & 0.081 & 0.797 \\
37 & XGB & FS6 & NONE & 0.041 & 0.080 & 0.799 \\
28 & XGB & FS1 & NONE & 0.029 & 0.077 & 0.758 \\
40 & XGB & FS4 & NONE & 0.039 & 0.072 & 0.809 \\
55 & LGBM & FS6 & NONE & 0.026 & 0.071 & 0.693 \\
39 & XGB & FS1 & NONE & 0.036 & 0.069 & 0.806 \\
45 & XGB & FS3 & NONE & 0.035 & 0.067 & 0.797 \\
8 & LR & FS2 & NONE & 0.027 & 0.066 & 0.726 \\
43 & XGB & FS1 & NONE & 0.030 & 0.064 & 0.837 \\
134 & RF & FS6 & NONE & 0.033 & 0.063 & 0.805 \\
68 & LGBM & FS1 & NONE & 0.022 & 0.061 & 0.737 \\
25 & XGB & FS4 & NONE & 0.029 & 0.057 & 0.782 \\
119 & ENS & STACK & NONE & 0.061 & 0.051 & 0.811 \\
62 & LGBM & FS4 & NONE & 0.029 & 0.050 & 0.781 \\
57 & LGBM & FS3 & NONE & 0.021 & 0.048 & 0.707 \\
35 & XGB & FS4 & NONE & 0.044 & 0.046 & 0.817 \\
58 & LGBM & FS1 & NONE & 0.026 & 0.043 & 0.770 \\
51 & LGBM & FS6 & NONE & 0.027 & 0.040 & 0.791 \\
143 & ENS & STACK & NONE & 0.094 & 0.039 & 0.833 \\
111 & XGB & FS4 & NONE & 0.070 & 0.039 & 0.829 \\
144 & ENS & STACK & NONE & 0.094 & 0.037 & 0.833 \\
44 & XGB & FS3 & NONE & 0.030 & 0.034 & 0.827 \\
48 & LGBM & FS3 & NONE & 0.016 & 0.033 & 0.692 \\
54 & LGBM & FS1 & NONE & 0.019 & 0.031 & 0.741 \\
103 & XGB & FS4 & NONE & 0.036 & 0.031 & 0.808 \\
63 & LGBM & FS4 & NONE & 0.019 & 0.025 & 0.650 \\
33 & XGB & FS3 & NONE & 0.031 & 0.024 & 0.828 \\
66 & LGBM & FS4 & NONE & 0.019 & 0.024 & 0.674 \\
56 & LGBM & FS3 & NONE & 0.020 & 0.024 & 0.639 \\
61 & LGBM & FS1 & NONE & 0.019 & 0.023 & 0.726 \\
108 & LGBM & FS3 & NONE & 0.019 & 0.023 & 0.636 \\
106 & LGBM & FS3 & NONE & 0.019 & 0.023 & 0.636 \\
105 & LGBM & FS3 & NONE & 0.019 & 0.023 & 0.636 \\
70 & LGBM & FS1 & NONE & 0.017 & 0.022 & 0.649 \\
59 & LGBM & FS1 & NONE & 0.023 & 0.020 & 0.731 \\
107 & LGBM & FS3 & NONE & 0.019 & 0.019 & 0.636 \\
114 & LGBM & FS3 & NONE & 0.019 & 0.017 & 0.636 \\
99 & RF & FS3 & NONE & 0.000 & 0.000 & 0.000 \\
98 & RF & FS3 & NONE & 0.000 & 0.000 & 0.000 \\
72 & RF & FS6 & NONE & 0.000 & 0.000 & 0.000 \\
96 & XGB & FS1 & NONE & 0.000 & 0.000 & 0.000 \\
95 & XGB & FS1 & NONE & 0.000 & 0.000 & 0.000 \\
76 & RF & FS6 & NONE & 0.000 & 0.000 & 0.000 \\
93 & LGBM & FS1 & NONE & 0.000 & 0.000 & 0.000 \\
86 & XGB & FS4 & NONE & 0.000 & 0.000 & 0.000 \\
92 & LGBM & FS1 & NONE & 0.000 & 0.000 & 0.000 \\
77 & RF & FS3 & NONE & 0.000 & 0.000 & 0.000 \\
90 & LGBM & FS3 & NONE & 0.000 & 0.000 & 0.000 \\
89 & LGBM & FS3 & NONE & 0.000 & 0.000 & 0.000 \\
78 & RF & FS1 & NONE & 0.000 & 0.000 & 0.000 \\
87 & XGB & FS4 & NONE & 0.000 & 0.000 & 0.000 \\
129 & LGBM & FS3 & PLT & 0.063 & 0.000 & 0.762 \\
\end{longtable}
}

\end{document}